\documentstyle[aps,twocolumn,pra,epsf,eqsecnum]{revtex}
\begin{document}
\title{Cooling atoms in an optical trap by selective parametric excitation}
\author{N. Poli, R.J. Brecha\thanks{permanent address, Physics Dept., University 
of Dayton, 
Dayton, OH.}, G. Roati\thanks{also at Dipartimento di Fisica, Universit\`a di
    Trento, I-38050 Povo (Tn), Italy.}, and G. Modugno}
\address{INFM, LENS, and Dipartimento di Fisica, Universit\'a di Firenze, Largo 
E. Fermi 2, 50125 Firenze, Italy}

\date{\today}
\draft
\maketitle

\begin{abstract}
  We demonstrate the possibility of energy-selective removal of cold
  atoms from a tight optical trap by means of parametric excitation of the
  trap vibrational modes. Taking advantage of the anharmonicity of the
  trap potential, we selectively remove the most energetic trapped
  atoms or excite those at the bottom of the trap by tuning the
  parametric modulation frequency. This process, which had been
  previously identified  as a possible source of heating, also appears
  to be  a robust way for forcing evaporative cooling in
  anharmonic traps.
\end{abstract}
\narrowtext 

Parametric excitation of atomic motion due to laser intensity
fluctuations has been recently indicated as one of the major sources
of heating in far-off resonance optical traps (FORTs)
\cite{savard97,kimble00}, and as a limitation in achieving very long
trapping times \cite{ohara99,engler00}.  In this Letter we show that
parametric excitation in a FORT can also lead to a cooling of the
trapped sample, since the anharmonicity of the potential allows a
selective removal of the most energetic atoms alone.\par The
excitation process has typically been modeled assuming a harmonic FORT
potential \cite{savard97,kimble00}. For a modulation of the potential
depth at twice the trap vibrational frequencies, an exponential
heating of the trapped sample, accompanied by trap losses, has been
predicted.  The main features of the parametric heating process
predicted by the harmonic theory have been experimentally observed
\cite{friebel98,vuletic98,ye99}. Going beyond the basic trap model, it
has been possible to further refine theoretical calculations through
modeling of the anharmonic trapping potential used in actual
experiments\cite{jauregui01}. One of the consequences of the
anharmonicity is a variation of the trap frequencies with vibrational
energy, which implies that a modulation at a single frequency will be
resonant with a few trap levels only. \par
We present here a detailed study of the temperature of a sample of
cold atoms in a one-dimensional (1D) lattice FORT, in which the
anharmonicity of the potential is exploited to perform
energy-selective excitation and removal of the atoms, resulting in
either heating or cooling of the remaining trapped sample. In
particular, the cooling process appears to be a robust and relatively
efficient way to lower the atomic temperature via forced
evaporation.\par
The trap is created with a beam of single-mode radiation
generated by a Ti:sapphire laser at $\lambda_T$=787~nm, red-detuned by
approximately 18 nm from the D-lines of potassium.  The laser beam,
after passing through an acousto-optic modulator (AOM), is focused to
$w$=90~$\mu$m within a two-lens telescope, and then retroreflected to
form a vertical standing-wave. The focusing is chosen to yield a large Rayleigh length,
$z_R$=3~cm, which results in an almost uniform lattice depth along the
vertical direction. Therefore, the optical potential close to the beam
waist has the form
\begin{equation}
U(r,z)=-U_0\,\cos^2(kz)\,\exp\Big(-\frac{2r^2}{w^2}\Big)\,,
\label{pot}
\end{equation}
where $k=2\pi/\lambda_T$.  About 10$^7$ atoms are prepared in a
magneto-optical trap (MOT) at a temperature of 60~$\mu$K, and
typically 5\% of them are transferred to the FORT thanks to a
compression procedure described in \cite{roati01}.  After loading the
atoms in the FORT, the MOT beams are extinguished by means of AOMs and
mechanical shutters, and the atoms are left to evolve in the presence
of the trap light only. The density and the radial temperature of the
trapped atoms are determined using absorption imaging of the atomic
cloud after 1~ms of expansion following release from the FORT. The
typical vertical size of the cloud of 700~$\mu$m indicates that
approximately 1800 lattice sites are occupied by an average of 300
atoms each. The axial temperature of the atoms is measured with a
time-of-flight (TOF) technique, by absorption of a sheet of light
passing about 1.5~cm below the FORT. The temperature of the trapped
atoms, after a 100-ms equilibration phase following loading, stays
constant during storage in the FORT \cite{roati01}, and the radial and
axial temperatures are found to be equal within the uncertainty.  The
lifetime of the FORT is limited to about 1.2~s mainly by collisions
with background gas.\par The harmonic approximations for the radial
and axial frequencies of the single lattice site are measured with a
parametric excitation technique, as described below. Typical values
for these quantities are $\omega_r$=2$\pi\times$1.3~kHz,
$\omega_a$=2$\pi\times$700~kHz, which for the typical temperature
$T$=80~$\mu$K lead to a density of 10$^{12}$~cm$^{-3}$. From
the measured value of $\omega_a$ it is possible to estimate the trap
depth as $U_0=M\lambda_T^2\omega_a^2/8\pi^2$, which corresponds to
$U_0$=700~$\mu$K. \par
\begin{figure}
\begin{center}
\leavevmode\epsfxsize=9cm
    \epsfbox{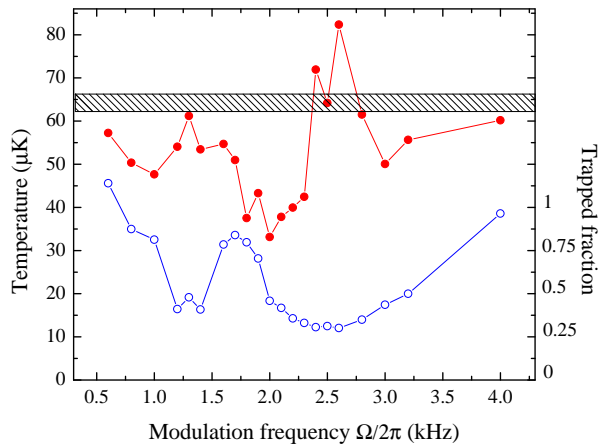}
\end{center}
\caption{Spectrum of the resonances in the trap losses (open circles)
  and in the radial temperature (solid circles) following a $T_{ex}$=50~ms,
  $\epsilon$=30\% excitation of the radial vibration. The shaded area
  indicates the initial temperature range, and represents the
  uncertainty in temperature. }
\label{exradial}
\end{figure}
\begin{figure}
\begin{center}
\leavevmode\epsfxsize=8cm
    \epsfbox{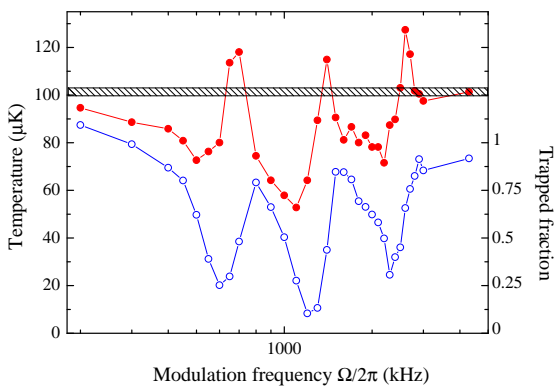}
\end{center}
\caption{Spectrum of the resonances in the trap losses (open circles)
  and in the radial temperature (solid circles) following a  $T_{ex}$=200~ms,
  $\epsilon$=16\% excitation of the axial vibration. }
\label{exaxial}
\end{figure}

We apply the parametric excitation via the control AOM as a small
sinusoidal modulation on the power of the trapping laser, typically
200~ms after loading the FORT.  To measure the spectrum of excitation
we have repeated at different modulation frequencies a procedure
composed of an excitation for a period $T_{ex}$ at fixed modulation
amplitude $\epsilon$ and frequency $\Omega$, followed by detection of
the number of atoms and radial temperature. For both radial and axial
excitation, as shown in Fig.~\ref{exradial} and Fig.~\ref{exaxial}
respectively, we see large resonances in the loss of atoms,
accompanied by dispersive resonances in the temperature of the atoms
left in the trap.  The resonances in the trap losses have been
observed previously in optical lattices
\cite{friebel98,jauregui01,roati01}, and their presence is expected
even if we approximate the FORT as a harmonic potential: when $\Omega$
is close to $2\,\omega_a$ or $2\,\omega_r$ (or to their subharmonics)
the atoms are excited upwards along the ladder of energy levels
\cite{savard97}, and eventually lost from the trap.  Note that for the
radial excitation two resonances at about $\omega_r$ and $2\,\omega_r$
are present, as expected for a harmonic trap, while in the axial case
there is an additional resonance close to $3.6\,\omega_a$, which is a
clear signature of the anharmonicity of the trap
\cite{jauregui01}.\par
\begin{figure}
\begin{center}
\leavevmode\epsfxsize=8cm
    \epsfbox{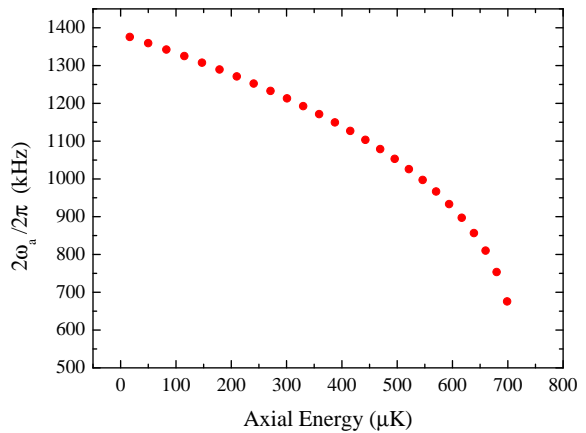}
\end{center}
\caption{Calculated spectrum of the axial excitation frequency for a 1D
    optical lattice with $U_0$=700~$\mu$K.}
\label{model}
\end{figure}

 On the other hand, the observed shape
of the resonances in the temperature of the remaining trapped atoms are not
expected for a harmonic trap \cite{savard97,kimble00}. Their origin
must be related to the anharmonicity of the potential, which implies
that the vibrational frequency changes with atomic energy. In
Fig.~\ref{model} we show the excitation frequency $2\,\omega_a$
calculated for a perfectly 1D sinusoidal lattice: while the atoms at
the bottom of the trap are expected to oscillate at the harmonic
vibrational frequency, the most energetic ones have a much lower
frequency, because of the decreased curvature of the potential.  A
similar behavior is also expected for the radial oscillation frequency
because of the gaussian shape of the radial potential. Therefore, when
the modulation frequency is tuned to the red of $2\,\omega_r$,
$2\,\omega_a$ or their harmonics and subharmonics, it is mainly the
most energetic atoms that are parametrically excited and ejected from
the trap, resulting in a reduction of the mean energy of the remaining
atoms. On the contrary, resonant modulation frequencies excite
primarily atoms at the bottom of the trap, resulting in a net increase
of the mean energy. We therefore identify the quantities
$2\,\omega_r$, $2\,\omega_a$ with the position of the main resonances
in the temperature spectra of Fig.~\ref{exradial} and
Fig.~\ref{exaxial} leading to a heating of the trapped sample.\par
These observations are an evidence of theoretical predictions that the
peak in the resonance of atomic losses in the trap occurs at a
slightly lower frequency than the actual harmonic frequencies
\cite{jauregui01}. This kind of information is very useful, since the
frequencies measured from the trap loss resonances are used to
estimate quantities such as the number density or the phase-space
density of atoms in an optical lattice \cite{friebel98,roati01}. \par
Note that the temperature measurements in Fig.~\ref{exradial} and
Fig.~\ref{exaxial} have been performed on the radial degree of freedom
only, since the TOF detection of the axial temperature is strongly
perturbed by the atoms excited out of the trap as long as 100~ms
before release. In all the measurement we report here we have
extracted the radial temperature from the width of the atomic
distribution, which was always well fitted by a gaussian, assuming
thermal equilibrium.\par
In the following we restrict our attention to the excitation of the
axial vibrational mode, which is the most energetic and therefore
requires a smaller perturbation of the trapping potential
\cite{axial}, making modeling easier.  We also suspect that the large
broadening of the radial resonances in Fig.~\ref{exradial} might come
from an ellipticity of the laser beam that makes $\omega_r$ vary in
the radial plane, masking interesting characteristics of the
resonances' shape.\par
To analyze in more detail the axial resonances we need to refine the
simple 1D model for the dispersion of the axial vibrational
frequencies presented in Fig.~\ref{model}. Indeed, when considering
the actual potential of Eq.~\ref{pot} one would expect that the axial
and radial degrees of freedom are mixed by the anharmonicity. The main
consequence of the mixing is that  $\omega_a$ depends also on
the total energy, and not only on the axial energy. Since
the axial and radial oscillation periods differ by a factor of about
500, it is possible to define a local axial frequency which varies
during the radial motion of the atoms \cite{heidelberg}
\begin{equation}
 \omega_a(r)=2\pi\sqrt{2U_0\exp(-2r^2/w^2)/M 
\lambda_T^{2}}\,.
\label{omegavar}
\end{equation}
As a result, the discrete excitation frequencies shown in
Fig.~\ref{model} for the 1D case become bands in 3D. The width of
these levels is
expected to be negligible for atoms at the bottom of the trap, which
experience an harmonic potential, and to increase for increasing total
energy, because of the longer time spent by the atoms in the
anharmonic part of the potential.  Thus, while the energetic atoms can
be excited by a large range of modulation frequencies on the red of
each parametric resonance, the ones at the bottom of the potential can
be excited only by a narrow range of frequencies close to $\omega_a$
and its harmonics, as confirmed by the temperature spectrum in
Fig.~\ref{exaxial}.\par
\begin{figure}
\begin{center}
\leavevmode\epsfxsize=8cm
    \epsfbox{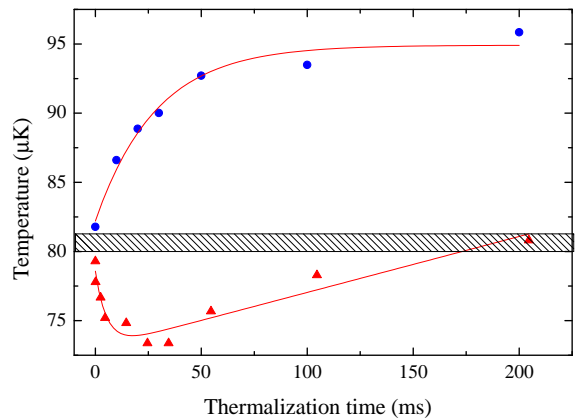}
\end{center}
\caption{Equilibration of the radial degree of freedom after an 
  excitation at 1.8$\omega_a$ (triangles), with $T_{ex}$=2~ms and
  $\epsilon$=12\%, and at 2$\omega_a$ (circles), with $T_{ex}$=10~ms
  and $\epsilon$=12\%. The solid curves represent the best fit with an
  exponential equilibration, plus a constant heating rate for the
  lower set of data.}
\label{therm}
\end{figure}

The role of anharmonicity in the excitation of atoms
having different energies is confirmed by a measure of the
equilibration time for the radial degree of freedom, following a short
axial excitation period, as reported in Fig.~\ref{therm}.  When
$\Omega$ is tuned to the red of $2\,\omega_a$, the equilibration time
is of the order of 5~ms, almost an order of magnitude shorter than the
elastic collisional time expected for our sample \cite{roati01}.  This
observation indicates that anharmonic cross-dimensional mixing can
actually be the dominant process in this case, as expected for a
perturbation of the high-lying trap states. If $\Omega$ is tuned to
$2\,\omega_a$ the two degrees of freedom thermalize on a longer
timescale of 20-50~ms, which is more likely to be determined by the
rate of interatomic collisions alone. In fact the thermalization
time shows an almost linear dependence on the atomic density, as
already observed in Refs.\cite{vuletic99,roati01}. In accordance with our
heuristic arguments, it appears that anharmonic mixing is not
important for excitation of atoms at the bottom of the trap. \par
\begin{figure}
\begin{center}
\leavevmode\epsfxsize=8cm
    \epsfbox{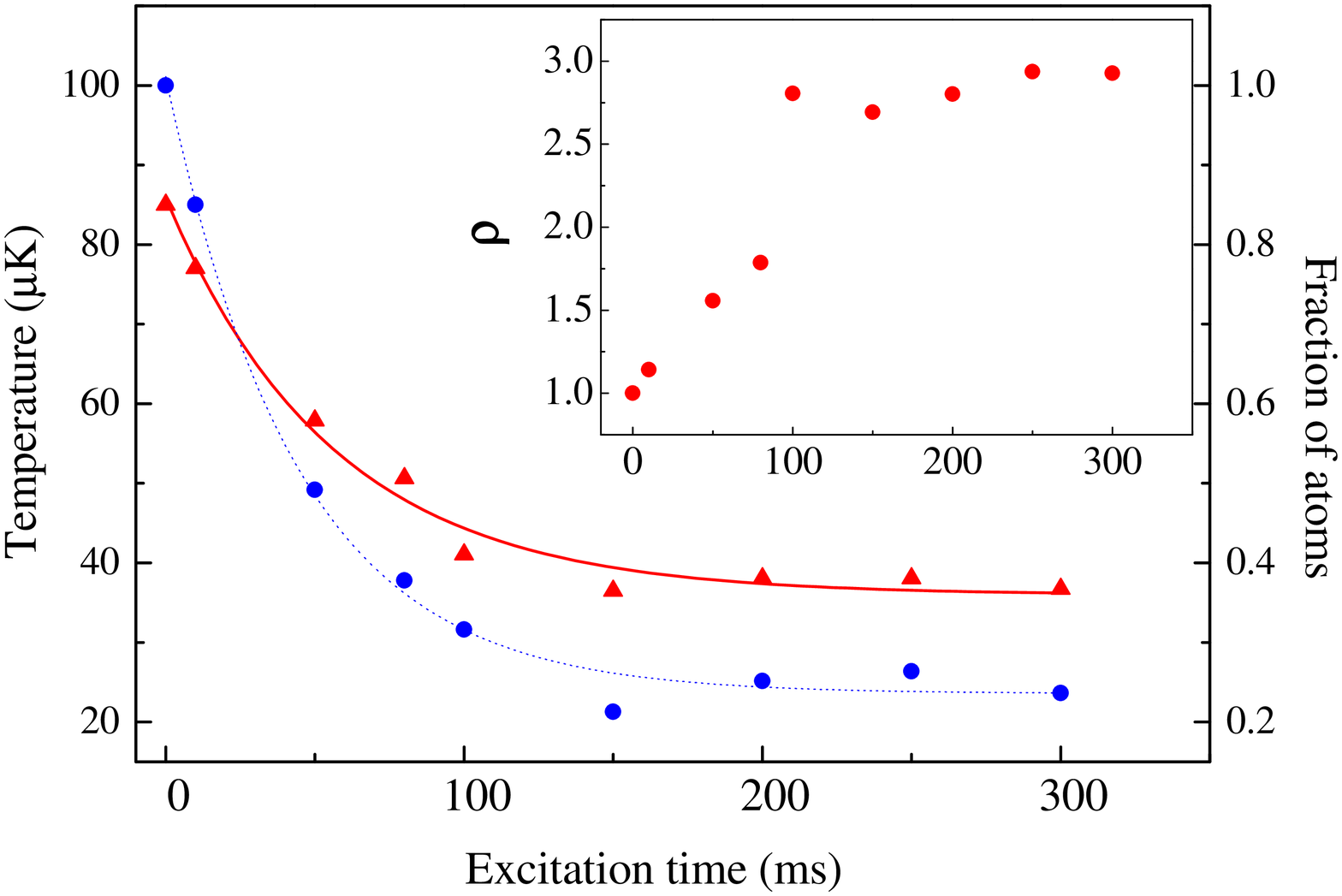}
\end{center}
\caption{Decrease of the radial temperature (triangles) and of the
  number of atoms (circles), during a parametric axial excitation with
  $\epsilon$=12\% and $\Omega$=1.6 $\omega_a$. The increase in
  relative phase-space density is shown in the inset}
\label{evaporation}
\end{figure}

We have investigated the time evolution of losses and temperature for
modulation frequencies red-detuned from $2\,\omega_a$, to get an
indication of the efficiency of the selective parametric excitation in
cooling the atoms in the FORT. The evolution reported in
Fig.~\ref{evaporation} was obtained with $\epsilon$=12\%,
$\Omega=1.6\,\omega_a$, and $T_{ex}$ variable between 5 and 300~ms;
the atom number and temperature were detected after a 20-ms
thermalization period. The general features of the process discussed
above are thus confirmed: at the beginning of the excitation phase the
energetic atoms are very efficiently removed from the trap, and very
likely the selection is performed on the total (3D) energy (the
reduction of the radial temperature following axial excitation is very
fast). We think that elastic collisions are responsible for a
continuous replenishing of the high-lying states of the trap, giving
rise to the observed exponential decrease of both atom number and
temperature on a timescale much longer than 5~ms.
Thus we can consider this process to be one of evaporative cooling,
forced by the parametric excitation. As shown in
Fig.~\ref{evaporation}, after 150~ms the atom number and temperature
reach a steady state value, which is conserved even after a further
150~ms period of excitation. The features just described are very
interesting from the perspective of using this technique to cool the
atoms in the trap, since an excitation at the proper frequency
provides a fast removal of the energetic atoms without heating the
coldest ones. After switching-off the parametric modulation at
$t$=300~ms we have measured a temperature increase at a rate of
34(12)~$\mu$Ks$^{-1}$ (compatible with that expected for photon
scattering), which indicates that the minimum temperature reached in
the cooling phase is probably not limited by the excitation process
itself.\par
From the two sets of data in
Fig.~\ref{evaporation} we extract the evolution of the relative phase space
density $\rho$ reported the inset, assuming thermal equilibrium
between radial and axial motions. Our experimental value for the parameter
which describes the overall efficiency of an evaporative cooling
process \cite{ketterle} is
\begin{equation}
\gamma_{tot}=\frac{N_f \rho_f}{N_i \rho_i}=0.7(1)\,.
\end{equation}
It can be compared to the efficiency of a variety of evaporative
techniques in optical and magnetic traps, as reported in Ref.
\cite{ketterle}. To our knowledge the process we have investigated
here represents the only way of forcing evaporative cooling in a
standard optical trap \cite{wieman} without a reduction in the trap
depth, a technique investigated in Ref. \cite{adams95}.\par Our FORT
is actually not optimized for evaporation; indeed we are obliged by
the short lifetime of the trap to force the removal of atoms on a
timescale not much larger than the elastic collisional time.  Further
investigations in a better vacuum environment, or at a larger atomic
density, can provide better tests of the efficiency of this cooling
technique.  Indeed, with a longer time for evaporation one could try
to use a modulation frequency on the red of all the axial resonances
($\Omega<\omega_a$), to excite only the very energetic atoms and thus
increase the cooling efficiency. Also, one could try to investigate
the effect of a sweep of $\Omega$ towards resonance during cooling, in
an analogous way to what is usually performed in magnetic traps.\par
In conclusion, we have reported the observation of an energy-selective
removal of atoms from a FORT, as a consequence of parametric
excitation of anharmonic vibrational modes. In addition, we have shown
that the energy selectivity can be exploited to cool the trapped
sample.  Since this effect arises due to the anharmonicity of the
trapping potential, it should be observable in most actual traps for
cold particles. The removal mechanism should work independently of the
internal structure of the trapped particles (atoms, molecules or
ions), as it is based on the excitation of their external degrees of
freedom.  The observations reported here should provide the impetus
for further investigations and theoretical modeling necessary for
characterizing the ultimate limits and efficiency of this cooling
technique; however, it already appears interesting for a fast cooling
of an atomic sample in a tight FORT.\par We thank M. Inguscio for
useful suggestions and for careful reading of the manuscript. We also
acknowledge useful discussions with R. J\'auregui, and we thank Scuola
Normale Superiore, Pisa for the loan of the trapping laser. This work
was supported by the ECC under the Contract HPRICT1999-00111, and by
MURST under the PRIN 2000 Program.  R.B. was partially supported by
ASI.

\begin{thebibliography}{999}
\bibitem{savard97} T. A. Savard, K. M. O'Hara, and J. E. Thomas, 
Phys. Rev. A{\bf 56}, R1095 (1997); M. E. Gehm, K. M. O'Hara, T. A. Savard, and J. E. Thomas, Phys. Rev. A {\bf 58}, 3914 (1998).

\bibitem{kimble00} C. W. Gardiner, J. Ye, H. C. Nagerl, and
  H. J. Kimble, Phys. Rev. A {\bf 61}, 045801 (2000).

\bibitem{ohara99} K. M. O'Hara, S. R. Granade, M. E. Gehm, and
  T. A. Savard, Phys. Rev. Lett. {\bf 82}, 4204 (1999).

\bibitem{engler00} H. Engler, T. Weber, M. Mudrich, R. Grimm, and M. Weidem\"uller, Phys. Rev. A {\bf 62}, 031402 (2000).

\bibitem{friebel98} S. Friebel, C. D'Andrea, J. Walz, M. Weitz, and
  T. W. H\"ansch, Phys. Rev. A {\bf 57}, R20 (1998).

\bibitem{vuletic98} V. Vuletic, C. Chin, A. J. Kerman, and S. Chu,
  Phys. Rev. Lett. {\bf 81}, 26(1998).

\bibitem{ye99}J. Ye, D. W. Vernooy, and H. J. Kimble, Phys. Rev. Lett. {\bf 83}, 4987 (1999).

\bibitem{jauregui01} R. J\'auregui, N. Poli, G. Roati, and G. Modugno,
  submitted to Phys. Rev. A; arXiv: physics/0103046.

\bibitem{roati01} G. Roati, W. Jastrzebski, A. Simoni, G. Modugno, and M.
Inguscio, to appear in Phys. Rev. A; arXiv: physics/0010065.

\bibitem{axial} A general result of the theory presented in \cite{savard97} is that the excitation rate is proportional to $\epsilon^2\omega^2$. 

\bibitem{heidelberg} T. Els\"asser, diploma thesis, University of
  Heidelberg (2000).

\bibitem{vuletic99} V. Vuletic, A. J. Kerman, C. Chin, and S. Chu, Phys. Rev. Lett. {\bf 82}, 1406 (1999).

\bibitem{ketterle} W. Ketterle and N. J. van Druten,
  Adv. At. Mol. Opt. Phys. {\bf 37} 181 (1996).

\bibitem{wieman} An exception is represented by the optical trap
  investigated in K.L. Corwin, S. J. M. Kuppens, D. Cho, and C. E. Wieman, Phys. Rev. Lett. {\bf 83}, 1311 (1999).

\bibitem{adams95} C. S. Adams, H. J. Lee, N. Davidson, M. Kasevich,
  and S. Chu, Phys. Rev. Lett. {\bf 74}, 3577 (1995).

\end{thebibliography}
\end{document}